%% file: ay12.tex
\documentstyle[preprint,aps]{revtex}
\begin{document}

\def\bsig{\mbox{\boldmath{$\sigma$}}}
\def\bl{{\bf l}}
\def\haf{\textstyle{1\over2}}

\title{The $A_y$ Puzzle and the Nuclear Force}

\author{D. H\"uber and J.L. Friar}

\address{Theoretical Division, Los Alamos National Laboratory, 
M.S. B283, Los Alamos, NM 87545, USA}

\maketitle

\begin{abstract}
The nucleon-deuteron analyzing power $A_y$ in elastic nucleon-deuteron
scattering poses a longstanding puzzle. At energies $E_{lab}$ below
approximately 30 MeV $A_y$ cannot be described by any realistic NN force. The
inclusion of existing three-nucleon forces does not improve the situation.
Because of recent questions about the $^3\!P_J$ NN phases, we examine whether
reasonable changes in the NN force can resolve the puzzle. In order to do this
we investigate the effect on the $^3\!P_J$ waves produced by changes 
in different
parts of the potential (viz., the central force, tensor force, etc.), 
as well as
on the 2-body observables and on $A_y$. We find that it is not possible with
reasonable changes in the NN potential to increase the 3-body $A_y$ and at the
same time to keep the 2-body observables unchanged. We therefore conclude that
the $A_y$ puzzle is likely to be solved by new three-nucleon forces, such as
those of spin-orbit type, which have not yet been taken into account.
\end{abstract}
\newpage

\section{Introduction}
\label{intro}

The so-called $A_y$ puzzle is a longstanding problem in elastic
nucleon-deuteron (Nd) scattering. Since it was first possible to perform
rigorous nd scattering calculations\cite{cornelius} it has been known that the
nucleon vector analyzing power $A_y$ cannot be described by any realistic
nucleon-nucleon (NN) force at energies below $\approx 30$ MeV. The same is true
for the deuteron vector analyzing power $i T_{11}$, whereas the deuteron tensor
analyzing powers and the differential cross section, for example, can be
described very well. Thus one should speak of a vector analyzing power puzzle.
But because this problem is known in the literature as the $A_y$ puzzle we will
stick to that name. The puzzle remains after the introduction of the latest
generation of (nearly) phase-equivalent NN forces\cite{report}. Since it is now
possible to calculate elastic proton-deuteron (pd) scattering below the breakup
threshold\cite{pisa} as well, we know that the same problem exists there, too.

A first attempt to solve the $A_y$ puzzle was made in a purely phenomenological
study\cite{aypaper}. Because $A_y$ is mainly sensitive to the NN $^3\!P_J$ phase
shifts\cite{aypaper}, the potentials for those partial waves were multiplied by
strength factors, keeping the low-energy observables in reasonable agreement
with the 2-body data, while at the same time increasing $A_y$ predictions
towards the experimental data. This could be achieved by introducing large
charge-independence breaking (CIB) and charge-symmetry breaking (CSB) into the
NN potential. Though such a large CIB and CSB is certainly unphysical, this 
study
suggested that there might be some room for changes in the NN potential.

A similar approach was adopted by Ref.\cite{wittor} (see also \cite{werner} and
\cite{twk}), 
where it was claimed that
there exists some room for changes in the $^3\!P_J$ phase shifts at lower 
energies. The hope was
that one could find modified $^3\!P_J$ phase shifts that describe the NN data as
accurately as the phase shifts that result from the latest phase-shift
analyses\cite{Nijm} \cite{arndt} and at the same time increase $A_y$.

Another possible solution to the $A_y$ puzzle is a three-nucleon force (3NF).
In\cite{3nf} it became possible for the first time to incorporate a 3NF into
Faddeev calculations above the breakup threshold. Since then all available 
3NFs
have been tried\cite{94aypaper} \cite{report} \cite{pisa} \cite{pisa3nf1}
\cite{pisa3nf2}, but all of them
either produce no significant effect on $A_y$ or slightly worsen the situation.
Existing 3NF models typically contain only those terms  believed most important
and least complicated, so the final word on such models has not yet been 
spoken. We
should remember that at the time these models were developed it was not 
possible
to test them in any calculation. Thus it might very well be true that the
available
forces are missing terms that are essential for the vector analyzing powers.

The aim of this study is to determine which improvements in the $A_y$ problem
are possible by changes in the NN potential. A critical discussion of options is
given in section \ref{Op}. In our calculations we make use of the AV18 
potential\cite{AV18}, which is introduced in section \ref{NN}. The changes 
we apply to the
NN potential are described in section \ref{change}. In section \ref{size} we
discuss the size of the changes in the potential that are necessary to keep the
2-body observables unchanged and at the same time increase $A_y$. Section
\ref{opep} deals with the special role of the one-pion-exchange potential
(OPEP), which gives the longest-range part of the NN force. In section
\ref{phases} we discuss how the possible changes in the NN potential are
influenced by the requirement that the $^3\!P_J$ phase shifts should not be
changed. We also comment on Ref.\cite{werner} and discuss whether any changes in
the $^3\!P_J$ phase shifts might be able to improve $A_y$. The question of 
whether charge-independence and charge-symmetry breaking might be able
to improve the description of $A_y$ will be answered in section \ref{cibcsb}.
Finally we summarize and conclude in section \ref{sum}.

\section{Options}
\label{Op}

We briefly assess the available dynamical options, necessary assumptions, and 
uncertainties, and categorize them in the order they will be discussed below.

We assume the existence of a 2-body nuclear potential that is independent of the
energy. Although that potential is not uniquely defined (because of off-shell 
ambiguities), we assume that it is as
momentum independent as is allowed by the underlying strong-interaction theory
(this prescription is called ``minimal nonlocality'' in Ref. \cite{atoms}). 
Minimal nonlocality fixes the representation, eliminates off-shell ambiguities,
and specifies the form of the dominant part of the 
potential (such as OPEP). It corresponds most closely to the majority of 
potentials in existence today. Because off-shell freedom is equivalent to a
3NF, this prescription also defines the 3NF\cite{mec}\cite{3bf}. Such a 
2-body potential {\bf must} be momentum dependent because of special relativity,
but in low-momentum applications such as few-nucleon systems that dependence is
constrained by the nucleon mass $M$ (viz., the dependence is $\sim (p/M)$, which
is small) or by the large QCD scale of the same size as $M$. We do not expect 
such
momentum dependence to be a critical factor, unless it occurs in combinations
with the nucleon spin $\vec s$ such as $\vec  l \cdot \vec s$ (i.e., a
spin-orbit interaction). Although our approach is nonrelativistic, we note that 
nonlocal interactions (incorporating relativistic corrections in some cases)
that have been used in
studying $A_y$ produce virtually the same results as local ones.

We assume that any NN potential should produce a good fit to the NN data base.
This is our primary principle, and the criterion for rejecting options. The
quality of that fit does not have to be the best, but it should not be poor.
First-generation potentials (i.e., older potentials that do not fit the data
particularly well) do not differ significantly in $A_y$ calculations from newer
potentials that have a much better fit.

There is only one exception: the AV14 potential \cite{AV14} gives a much lower 
prediction for $A_y$ than all the other potentials \cite{report}. The reason
for this behavior is that the $^3\!P_J$ phases of AV14 deviate
strongly from those of all other potentials \cite{report} (which means that
AV14 does not fit the NN data base well enough).

The fact that all potentials with this one exception (due to differences 
in the phase-shift parameters) give essentially the same predictions for 
$A_y$ and $i T_{11}$ strongly argues that the $A_y$ puzzle is not a simple
problem of the off-shell behavior of the NN potential.
The NN potentials on the market (which were all tried on the vector
analyzing powers) vary from strictly local to strongly non-local 
and thus exhibit rather different off-shell behaviors. Experience
\cite{report} shows that the analyzing powers are insensitive to the
off-shell behavior of existing NN potentials. Thus the assumption above, that
it is sufficient that a potential give a reasonable fit to the NN data base,
appears justified within the context of ``minimal nonlocality''. We will 
comment more on this in section \ref{sum}.

We conceptually divide the potential into two parts: OPEP plus a shorter-range
part. Within this framework we have four possible options for improving the
description of $A_y$ without violating our primary principle.

The first option is to change OPEP. These changes could arise from changing the
pion-nucleon coupling constant, by modifying the virtual-pion spectral 
function,
by momentum-dependent modifications due to special relativity, and by vertex
modifications (i.e., form factors). The current status of the pion-nucleon
coupling constant is reviewed in Ref.\cite{vanc}. The bulk of the phase-shift
analyses (including the energy-dependent analyses) favor a common low value. The
potentials that we use all have this value. The pion spectral function is a
two-loop modification of the propagator and consequently is very tiny\cite{dpc}.
Form factors are a short-range modification, which is discussed below. The
effects of relativity have been examined in three-nucleon bound-state
calculations, where they are rather small, but no fully relativistic scattering
calculations have been performed. Isospin violation is already included in part
through the use of different charged- and neutral-pion masses. There is no
evidence for different charged- and neutral-pion-nucleon coupling
constants\cite{isospin} at the 1\% level.

The short-range interactions are parameterized using as many different forms as
there are potentials. The functional forms include Gaussian, Yukawa, and Fermi
functions, and combinations thereof. As we shall see, at low energies a single
parameter describes this interaction in the P-waves. At higher energies and in
S-waves one or two more may be needed. In effect only a few moments of the
potential are required by the NN data, and this is easy to impose on an
arbitrary functional form. If OPEP is fixed, this is the primary freedom.

Isospin violation has been suggested as a candidate for solving the $A_y$
puzzle. Because the {\bf same} problem exists for pd (which has no nn
interaction) as for nd scattering (which has no pp interaction), it very likely
cannot involve CSB (charge symmetry interchanges protons and neutrons). The bulk
of the charge dependence (CD) 
is already included via the different pion masses. What remains should
have short range and be rather small in P-waves. 
We will comment further on isospin violation later in the paper.

Finally, introducing a 3NF does not affect the two-nucleon problem. Although
those forces used to date have not helped to resolve the puzzle, there are
additional components of the two-pion-range 3NF that have a spin-orbit character
and have never been included in a calculation\cite{3bf,vk1}. In the early days 
of building force models it was conventional to ignore momentum-dependent forces
(and thus anything proportional to $\vec l$), because of the complexity.

\section{The NN potential}
\label{NN}

Since earlier attempts to resolve the $A_y$ puzzle (by multiplying the NN
potentials in the $^3\!P_J$ waves with strength factors\cite{aypaper}
\cite{report}) did not lead to satisfying results, we will pursue an alternative
approach that introduces more flexibility in changing the potential and thus
more possibilities for resolving the puzzle. We apply different strength factors
to different parts of the potential, thereby introducing additional freedom (in
the form of parameters); this enhances the possibility of finding a set of
parameters that leaves the 2-body observables unchanged and at the same time
increases the 3-body analyzing powers $A_y$ and $i T_{11}$ in the desired
manner. Our goal will be to relate changes in the NN force to changes in the np
and nd analyzing powers. We emphasize that we are not advocating large changes
in the potential that are unsupported by the NN data. Rather, our goal is to
gain insight into these relationships before drawing any conclusions.

For this purpose we have chosen the AV18 potential\cite{AV18}, which is well
suited to use in the study because of its structural simplicity. The AV18
potential is a semi-phenomenological potential with a one-pion-exchange tail.
This potential is structured around 18 spin-isospin-orbital operators, which are
multiplied by different radial functions. In addition to one-pion-exchange
components (whose form is well understood and not controversial), those radial
functions contain a parameterized short-range component. The operators
themselves are constructed from the vectors that are available, such as the
distance between the two nucleons $\vec r$, the total 2-body angular momentum
$\vec l=\vec l_1 + \vec l_2$ and the total 2-body spin $\vec s=\vec s_1 + \vec
s_2$. The only constraint on the operators that are constructed from these
vectors is that they must be scalars. The parameters in the radial functions
were fitted to the Nijmegen NN data base with a $\chi^2$ per datum of slightly
more than one. Thus, the AV18 potential represents a very general form of the NN
potential with a good description of the NN data, and this meets our
requirements.

Because we are interested in the $^3\!P_J$ waves we have to deal with only 5 of
the 18 operators in AV18: the operator 1, which gives the central force $V_C$,
$S_{12}$, which gives the tensor force $V_T$, $\vec l\cdot\vec s$, which gives
the spin-orbit force $V_{ls}$, ${\vec l}^2$, which gives $V_{l^2}$, and finally
$(\vec l\cdot\vec s)^2$, which gives $V_{(ls)^2}$.

\section{Changing the NN potential}
\label{change}

In order to study the sensitivity of the 3-body analyzing power $A_y$ to the
above-mentioned five parts of the NN force, we increase and decrease by 10\%
each of those five parts of the NN force in the $^3\!P_J$ waves,  without
introducing any additional CIB or CSB.
All other partial waves remain unchanged. At the same time we
study the effect of these changes on the 2-body observables. It is well known
that the only 2-body observable sensitive to changes in the $^3\!P_J$ waves is the
analyzing power. In order to distinguish it from the 3-body analyzing power
($A_y$) we will call it $A_2$. Table \ref{tab1} shows the effect of the changes
on $A_2$ and Table \ref{tab2} on $A_y$.

Tables \ref{tab1} and \ref{tab2} demonstrate that the effects of a 10\% increase
or decrease are roughly the same. Therefore we can conclude that (within roughly
10\% ) these changes in the potential have a linear effect on $A_2$ and $A_y$.
In other words, each change of $\pm1$\% in $V_C$ causes a change in $A_2$ of
roughly $\pm0.3$\% at low energies 
(regardless of the starting point for that change), as long
as the total deviation from the original AV18 potential is less than $\pm10$\%.
For larger deviations from the original AV18 potential this linearity is lost.

Next we note that the NN force components that have the largest effect on $A_2$
are the spin-orbit and tensor forces; the effect of the tensor force on $A_y$ is
considerably smaller than the effect of the spin-orbit force, but still fairly
large. This is what one expects, since a vector analyzing power is defined by
\begin{equation}
\label{eq1}
A\equiv {\sigma_\uparrow - \sigma_\downarrow \over 
\sigma_\uparrow + \sigma_\downarrow}\, ,
\end{equation}
where $\sigma_\uparrow$ and $\sigma_\downarrow$ denote the differential cross
section with the spin of the incoming nucleon (for $A_y$) or deuteron (for
$i T_{11}$) oriented normal to the scattering
plane. Intuitively, such an asymmetry is generated by those potential terms
(such as $V_{ls}$ and $V_T$) that depend on the spin direction. The terms $V_C$
and $V_{l^2}$ do not depend on the spin at all, while $V_{(ls)^2}$ has less
influence because it is small. If we set $V_{(ls)^2}$ to zero, $A_2$ decreases
only by 4.6\%.

The most important point here is that only the effect of a change in the tensor
force is significantly different in the 2-body and the 3-body analyzing power.
(We shall explain below why this is so.) This implies that, if we want to keep
the 2-body prediction unchanged but want to change the 3-body prediction, this
must come from a change in the NN tensor force. Changes in the other parts of
the NN potential are then needed in order to compensate in $A_2$ for the change 
in $V_T$.

We note that the AV18 prediction for the nd $A_y$ at $E_{lab}=3$
MeV\cite{data_nd} underestimates the data near the maximum by about 30\%. Also,
from Tables \ref{tab1} and \ref{tab2} we learn that the effect of a change in
the tensor force is larger in the 2-body system than in the 3-body system. This
means that we first have to {\bf decrease} $A_2$ and $A_y$ by increasing $V_T$
and then {\bf increase} the analyzing powers again by changes in the other terms
until $A_2$ resembles its original value. $A_y$ will then have a larger value
than before because the effect of the decrease by $V_T$ was less for $A_y$ than
for $A_2$. An analogous argument explains why the 3-nucleon binding energy
increases with decreasing tensor force, if the deuteron binding energy is kept
fixed. These changes in $A_2$ and $A_y$ require large changes in the NN
potential.

We next quantify those changes in the potential that are necessary, although
only a rough estimate is required. In order to achieve this let us consider the
effects of changes in the various terms of the NN potential at 1 MeV (for
specificity), as shown in Table \ref{tab1}. The requirement that the totality of
changes in the potential not affect $A_2$ leads to the equation
\begin{equation}
\label{eq2}
.3\ \delta_C-1.2\ \delta_T+1.7\ \delta_{ls}-.5\ \delta_{l^2}
+[.04\ \delta_{(ls)^2}]=0 \, .
\end{equation}
The quantities $\delta$ denote the change (in percent) in the corresponding term
of the potential. The factors in front of the $\delta$'s in Eq. (\ref{eq2}) 
mean,
for example, that a change in $V_C$ of 1\% leads to a change in $A_2$ of roughly
0.3\%. $V_{(ls)^2}$ has nearly no effect on $A_2$, but has 
a small one on $A_y$, and
this is indicated by the brackets. We will first neglect this term and then take
it into account later.

We based Eq. (\ref{eq2}) on the results in Table \ref{tab1} at $E_{lab}=1$ MeV
for two reasons. First, the results at $E_{lab}=10$ MeV are very similar to the
ones at 1 MeV. Second, we concentrate for the moment only on $A_y$ at 
$E_{lab}=3$
MeV. The 2-body t-matrix for 2-body energies from 2 MeV to $-\infty$ are
required for Faddeev calculations of nd scattering at $E_{lab}=3$ MeV, so that
we can neglect the higher 2-body energies for the moment. Clearly, this is a
very rough procedure. For a real solution of the $A_y$ problem one would have to
consider all 2- and 3-body energies. But because we only require a rough
estimate of the size of the necessary changes in the potential, this procedure 
is good enough for the moment.

The analogue of Eq. (\ref{eq2}) that we get from Table \ref{tab2} is
\begin{equation}
\label{eq3}
.4\ \delta_C-.6\ \delta_T+1.6\ \delta_{ls}-.6\ \delta_{l^2}
-[.1\ \delta_{(ls)^2}]=30 \, ,
\end{equation}
which corresponds to an increase in $A_y$ of 30\%. 

Solving Eqs. (\ref{eq2}) and (\ref{eq3}) for $\delta_T$ and $\delta_{ls}$ we
obtain
\begin{eqnarray}
\label{eq4}
\delta_T&=&-.22\ \delta_C+.24\ \delta_{l^2}+56.7\, , \nonumber \\
\delta_{ls}&=&-.33\ \delta_C+.47\ \delta_{l^2}+40 \, .
\end{eqnarray}
Obviously $\delta_T$ and $\delta_{ls}$ become large numbers if $\delta_C$ and
$\delta_{ls}$ are chosen to be reasonably small, or, vice versa, if we require
$\delta_T$ and $\delta_{ls}$ to become reasonably small, $\delta_C$ and
$\delta_{ls}$ must be chosen very large. The inclusion of $\delta_{(ls)^2}$ in
Eqs. (\ref{eq2}) and (\ref{eq3}) does not help much. In that case we get
\begin{eqnarray}
\label{eq5}
\delta_T&=&-.22\ \delta_C+.24\ \delta_{l^2}+.26\  \delta_{(ls)^2}
+56.7 \, , \nonumber \\
\delta_{ls}&=&-.33\ \delta_C+.47\ \delta_{l^2}+.16\  \delta_{(ls)^2}
+40 \, .
\end{eqnarray}

We give several possible solutions of Eqs. (\ref{eq2}) and (\ref{eq3}) in Table
\ref{tab3} together with the effect of these changes on $A_2$ and $A_y$. Though
for all cases listed in Table \ref{tab3} there is a substantial increase of
$A_2$, the increase in $A_y$ is roughly a factor 5 to 10 larger and always far
above the required 30\%. The reason that the solutions of Eqs. (\ref{eq2}) and
(\ref{eq3}) listed in Table \ref{tab3} are so far away from the required 0\% and
30\% changes (for $A_2$ and $A_y$), respectively, is that the parameters in
Eqs. (\ref{eq2}) and (\ref{eq3}) are based on small ( $<$10\%) changes in the
potential. For the larger changes in the potential that are obviously necessary,
the factors in Eqs. (\ref{eq2}) and (\ref{eq3}) are energy dependent and no 
longer constants. Thus with some fine tuning it might be possible to reduce the
effect on $A_2$ to an acceptable level and still maintain an increase in $A_y$ 
of
about 30\%. But would that be the solution for the $A_y$ problem? Unfortunately
not. There are several reasons why this cannot be a solution for the $A_y$
problem and, moreover, why solving the $A_y$ problem with such changes in the NN
potential is not possible. We shall discuss this in the following sections.

\section{The size of the required changes in the NN potential}
\label{size}

Table \ref{tab3} shows that each solution of Eqs. (\ref{eq2}) and (\ref{eq3})
requires quite remarkable changes in the several terms of the NN potential. For
each of the tabulated solutions at least one term of the NN potential has to be
changed by more than 35\%. Changes of up to 50\% are required. Other solutions 
of Eqs. (\ref{eq2}) and (\ref{eq3}) than those shown in Table \ref{tab3} would
obviously result in similarly large changes in the NN potential. As mentioned in
section \ref{change}, in order to satisfy Eqs. (\ref{eq2}) and (\ref{eq3}) a 
huge change in the tensor force is unavoidable. This is illustrated by the fact
that in the expressions for $\delta_T$ and $\delta_{ls}$ in Eqs. (\ref{eq4})
and (\ref{eq5}) the multipliers of the various $\delta$'s on the right-hand 
sides are smaller by a factor of 100 than the last summand on each right-hand
side.

Such large changes in the NN potential can be ruled out. Though the AV18
potential is a semi-phenomenological potential, the strengths of its various
terms are not free. The radial functions in the AV18 potential are fit to the NN
data. Moreover, OPEP has been properly implemented and plays a large role, as we
discuss below.

The different terms in the potential were multiplied by constant strength
factors. One might argue that more freedom results if one changes the shape of
the radial functions in the potential, as well. Unfortunately, this will not
help much. Because the radial function for the one-pion-exchange potential
(OPEP) is well known, such a change could only be made for the radial functions
associated with the short-range operators in the potential. But it is
demonstrated in Tables \ref{tab1}, \ref{tab2}, \ref{tab4} and \ref{tab5} (for an
explanation of Tables \ref{tab4} and \ref{tab5}, see below) that such a
modification will result in roughly the same changes for $A_2$ and $A_y$ and
will thus not be able to solve the $A_y$ problem.

In fact, as we shall see in the next section, the freedom to change the NN
potential is even much more tightly constrained.

\section{The one-pion-exchange potential}
\label{opep}

As already mentioned there is one piece of the phenomenological AV18 potential
that comes from an important and well-recognized physical process - the OPEP.
The OPEP makes up the longest-range part of this potential, whereas the
short-range part is phenomenological. The OPEP has a tensor and a (weaker)
central part. As we shall see below, the pion tail in the tensor force is the
reason why the tensor force has a significantly different effect on the 2-body
and the 3-body analyzing powers.

Let us first regard Tables \ref{tab4} and \ref{tab5}. These tables show the
different effects of the short- and long-range parts of the central and tensor
forces on the analyzing powers. We accomplish this by separating the long- and
short-range parts into the form
\begin{equation}
\label{eq6}
V=V^{SR}+V^{OPEP}\, ,
\end{equation}
where $V^{SR}$ stands for the short-range part and $V^{OPEP}$ for the pion tail
of the potential. The latter also includes the short-range regulator that
makes it finite at the origin. Choosing a different regulation scheme is
equivalent to changing the short-range part.

Tables \ref{tab4} and \ref{tab5} demonstrate that all the sensitivity of the
analyzing powers to the tensor force comes from the pion tail. The short-range
tensor force has nearly no effect on the analyzing powers.

For the central force the pion tail is much less important than the short-range
part, which is well-known. What is important here is that changes in all
short-range operators cause roughly the same change in the 2-body and 3-body
analyzing powers. The reason is that the 2-body and 3-body matrix elements of a
short-range operator are roughly proportional to each other, as demonstrated in
Tables \ref{tab1}, \ref{tab2}, \ref{tab4} and \ref{tab5}. Thus if a change of
the strength of a short-range operator causes a certain relative change in the
2-body matrix element, the 3-body matrix element will be changed by roughly the
same relative amount.

Thus we find that if we don't want to change the 2-body analyzing power but want
to increase the 3-body analyzing power we must modify the long-range force.
Because the only long-range NN force is the OPEP, this is the part that must be
changed. The required increase is of order 30\% to 50\% in order to get the 30\%
increase in $A_y$, and this is unreasonable. Thus, all the solutions for Eqs.
(\ref{eq2}) and (\ref{eq3}) are inconsistent with the OPEP and therefore out of
the question.

In summary, the only way to keep $A_2$ unchanged and simultaneously raise $A_y$
would be a very large strengthening of the one-pion-exchange potential, and 
this would not be credible. Thus a solution of the $A_y$ problem by such 
changes in the NN potential is not possible.

In the next section we shall also demonstrate the importance of OPEP for
the $^3\!P_J$ phase shifts.

\section{The $^3\!P_J$ phase shifts}
\label{phases}

Up to now we have looked at observables, but we have not checked what
effects the changes that we made in the NN potential have on the $^3\!P_J$ phase
shifts. Table \ref{tab6} gives an overview.

The first noticeable thing in Table \ref{tab6} is that each of the different
terms of the potential affects the $^3\!P_J$ partial waves in very different
ways. The tensor force has the largest effect on the phases, although it does
not dominate $A_2$. The tensor force changes all three phases in the same
direction, $^3\!P_0$ and $^3\!P_2$ by roughly the same amount and $^3\!P_1$ by about
half this amount. Because the $^3\!P_J$ phases influence $A_2$ in different ways,
the changes in the phases partially cancel each other in $A_2$.

The spin-orbit force changes essentially only the $^3\!P_2$ phase, and by roughly
half the amount that the tensor force changes this phase. Because there is no
cancellation from the other two phases, the effect of the spin-orbit force on
$A_2$ is larger than that of the tensor force. The other three terms have only a
small influence on the phases and on $A_2$; they primarily affect $^3\!P_2$. The
central force has a modest effect on $^3\!P_1$.

What happens if we require that our changes to the potential leave not only
$A_2$ unchanged, but also the $^3\!P_J$ phases? Obviously we will get three
additional equations to be fulfilled together with Eqs. (\ref{eq2}) and
(\ref{eq3}). These are
\begin{eqnarray}
\label{eq7}
-.04\ \delta_C + 1.4\ \delta_T - .1\ \delta_{ls} -.07\ \delta_{(ls)^2} 
&=& 0 \, ,\nonumber \\
.1\ \delta_C + .8\ \delta_T + .03\ \delta_{ls} + .01\ \delta_{(ls)^2} &=& 0\, ,
\\
-.2\ \delta_C + 1.4\ \delta_T + .7\ \delta_{ls} - .3\ \delta_{l^2}
-.2\ \delta_{(ls)^2} &=& 0\, . \nonumber 
\end{eqnarray}
Thus we now have 5 equations for 5 unknowns. The solution of these 5 equations
is unique and gives $\delta_C=177$, $\delta_T=-19$, $\delta_{ls}=141$,
$\delta_{l^2}=577$ and $\delta_{(ls)^2}=-681$. Changes of this order are totally
out of question, and therefore we can conclude that reasonable changes in the NN
potential cannot keep the 2-body phase shifts and observables unchanged, while
at the same time increasing the 3-body analyzing power by the amount required by
the data.

In the discussion above we left out the $^3\!F_2$ phase and the 
$^3\!P_2$-$^3\!F_2$
mixing parameter $\epsilon_2$, because they are less important for the
analyzing powers. These parameters are also changed by the
modifications we made to the different potential terms. Requiring these two
parameters to be unchanged as well would lead to two additional equations
besides the three Eqs. (\ref{eq7}), so that then we would have 7 equations
for 5 unknowns. Unless there would be a redundancy within these 7 equations
this set would have no solution. Therefore our conclusions would remain the
same even if we would consider $^3\!F_2$ and $\epsilon_2$, too.

Heretofore we have made two assumptions: that there is no CIB and CSB in the
$^3\!P_J$ waves and that the $^3\!P_J$ phase shifts should not deviate from the
results of the Nijmegen phase shift analysis (PSA) (with which the AV18
potential is commensurate). 

The second assumption was questioned very recently by the author 
of\cite{werner} (see also Table III of \cite{twk}). He shows that there is 
room for some changes in the $^3\!P_J$
phase shifts at lower energies due to the fact that there are not enough
NN data to determine the low-energy $^3\!P_J$ phases
uniquely. In fact, in \cite{werner} the Fermi-Yang ambiguities 
were rediscovered, which were first found for $\pi$N scattering \cite{FY-piN}.
If a set of two or more phase shifts shows sensitivity in only one observable,
Fermi and Yang discovered that there is a continuous ambiguity in the 
determination of those phase shifts by a single-energy analysis.
(If a second observable shows sensitivity,
the ambiguities become discrete.) That is exactly the situation we face:
the only 2-body observable showing strong sensitivity to the $^3\!P_J$ phases
at lower energies is the analyzing power $A_2$. In a low-energy approximation
it can be written in the form
\begin{equation}
\label{eq8}
A_2(\theta)=f(\theta)(-2\delta_{\,^3\!P_0}-3\delta_{\,^3\!P_1}
+5\delta_{\,^3\!P_2}+c)
\end{equation}
The constant $c$ includes the dependence of partial waves other than $^3\!P_J$,
which play only a minor role. It is obvious from Eq. (\ref{eq8}) that
any combination of the $^3\!P_J$ phases that leaves the sum
$(-2\delta_{^3\!P_0}-3\delta_{^3\!P_1}+5\delta_{^3\!P_2})$ unchanged will do
equally well in the description of $A_2$. 

Thus these ambiguities are clearly there, but do they give any freedom
for changes in the $^3\!P_J$ phase shifts as determined in the Nijmegen PSA? 
They do not, as we shall show next.

First of all one can argue that in a multi-energy phase-shift analysis the 
low-energy $^3\!P_J$ phases are not only determined by low-energy NN data but by
other constraints (continuity and analyticity), whereas the analysis done in
\cite{werner} is equivalent to a single-energy phase-shift analysis and lacks
these constraints. Thus even if it is possible to describe the NN data at a
single energy with several sets of different $^3\!P_J$ phases (as it is
clearly shown
in\cite{werner}), it is virtually certain that all but one are ruled out in a
multi-energy phase-shift analysis.

Indeed, this is the longstanding position of
the Nijmegen group\cite{vanc}\cite{timm}: 
phase-shift ambiguities at a given energy are
removed by performing a multi-energy analysis and by adding additional physics.

\begin{figure}
\vskip-2truecm
\vbox{\hsize=15truecm\epsfbox{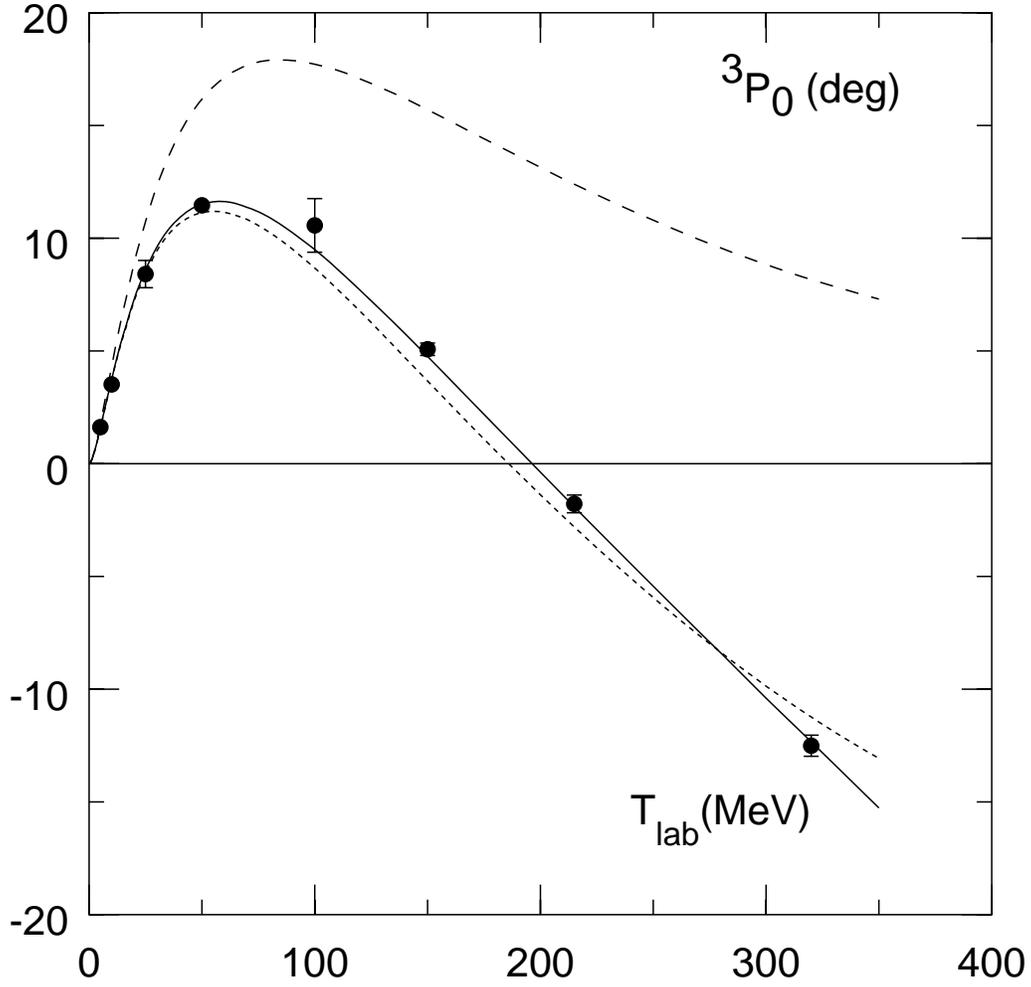}}
\vskip-5truecm
\caption{\label{fig4} The $^3\!P_0$ phase from the Nijmegen 
PSA\protect\cite{Nijm}. The dashed line is the prediction for OPEP only,
the dotted line is the result of the PSA with one parameter and the solid
line is the final result of the Nijmegen PSA with 3 parameters. The filled
circles denote results of single energy analysis.
}
\end{figure}

Physics in our case means OPEP. The inclusion of OPEP
indeed restricts the possible freedom for changes in the $^3\!P_J$ phases
drastically. This is demonstrated in Fig. 
\ref{fig4}, which is taken from Ref. \cite{Nijm}. Fig. \ref{fig4} shows
that the prediction for $^3\!P_0$ by OPEP alone already gives the correct
shape of this phase. Adding one more parameter into the PSA for shorter-range 
effects gives essentially the correct result. For a perfect $\chi^2$-fit only
two additional parameters are needed. In other words,
{\bf all} short-range effects in $^3\!P_0$ can be explained reasonably
well with one parameter only, which leaves very little room for
changes in this phase. 

Even if we assume that it is justified to modify the $^3\!P_J$ phases at low
energies within the limits given in\cite{werner}, one can show that this cannot
solve the $A_y$ problem. In Figures \ref{fig0} - \ref{fig2} we show the 
$^3\!P_J$ phase shifts as they change with energy. We have divided the phase
shifts by $E^{3/2}$, because the threshold behavior for phase shifts at low
energies is given  by $\delta_l\propto k^{2l+1}$. From Figures \ref{fig0} -
\ref{fig1} we see that $^3\!P_0$ and $^3\!P_1$ start to deviate from the
threshold behavior at 1 MeV, although not very strongly. The $^3\!P_2$ phase on
the other hand exhibits a nearly perfect threshold behavior up to 10 MeV (Figure
\ref{fig2}). This means that for $^3\!P_2$ at least, changes below 5 MeV must be
accompanied by corresponding changes up to 10 MeV and possibly higher. But at
those higher energies there is no room for changes in $^3\!P_2$\cite{werner};
thus it cannot be changed at lower energies either.

We next notice from Figures \ref{fig0} - \ref{fig2} that the short-range part of
the tensor force $V_T^{SR}$ has only a very small effect on the $^3\!P_J$ 
phases. This fact repeats our findings from Tables \ref{tab4} and \ref{tab5}
that the effect of the tensor force is almost exclusively in the 
one-pion-exchange part. Thus any major
changes in the tensor force (which are necessary for any improvement in the
$A_y$ problem, as we have seen above) must be made in the OPEP, which will not
accommodate a drastic change.

One might still argue that for small changes in the $^3\!P_J$ phases at low 
energies a change in the short-range tensor force might be sufficient. In order 
to show that this cannot be true, we have plotted in Figures \ref{fig0} -
\ref{fig2} the difference (dotted lines) between the phases for AV18 without the
short-range tensor force $V_T^{SR}$ and for the full AV18, again divided by
$E^{3/2}$. These lines are virtually constant for all three phases. This means
that the contribution of $V_T^{SR}$ to the $^3\!P_J$ phases exhibits a perfect
threshold behavior up to very high energies and thus $V_T^{SR}$ is essentially
determined by a single parameter for each of the $^3\!P_J$ waves. In other 
words,
a change in $V_T^{SR}$ leads necessarily to a change of the phase shifts for all
energies, and this change is proportional to $E^{3/2}$ up to very high energies.
It is impossible to change the phases at low energies via $V_T^{SR}$ without
disagreement with data at higher energies.

Let us go even one step further. Let us follow \cite{werner} and take $^3\!P_J$
phases that are modified at low energies in the spirit of\cite{werner}
as shown in Table \ref{tab6a}, 
and refit the AV18 potential to them. (Note that only $^3\!P_0$, $^3\!P_1$
and $^3\!P_2$ were modified at $E_{lab}=$1, 5 and 10 MeV, with the largest
modification required at 1 MeV and the smallest modification required at 
10 MeV.) Attempts\cite{bob} to fit
the potential to these modified $^3\!P_J$ phases were unsuccessful, unless the
pion-nucleon coupling was weakened. The reason for the necessity of weakening
the pion coupling was that Table \ref{tab6a} requires a weakening in $^3\!P_0$ 
by
5\% at the lowest energy, and this could only be achieved\cite{bob} by a weaker
pion-nucleon coupling. This is in perfect agreement with our findings above.
According to Table \ref{tab6} only a weaker tensor force can decrease $^3\!P_0$
significantly, and as we see in Table \ref{tab4} this requires a weaker OPEP.

It is also interesting to note from Table \ref{tab6a} that the attempt to fit
the phases that were modified at the lower energies led to changes in the 
phase-shift
predictions of the potential at all energies. The most dramatic changes in
the phase-shift prediction of the refitted potential (in comparison to the
original potential) are found at the higher energies for $^3\!F_2$. Most of the
predictions of the refitted potential for $^3\!F_2$ fall outside the error
bars of the Nijmegen PSA\cite{Nijm}. We based this judgment on the error
bars that are given in \cite{Nijm} for the pp phases; unfortunately Ref.
\cite{Nijm} gives no error bars for the np isovector phases. The refit
procedure for the potential was also based on these error bars\cite{bob}.

This shows that a refit of the AV18 potential to the energy-dependent modified
phases of Table \ref{tab6a} within the error bars for the phase shifts as
given in Ref. \cite{Nijm} is not possible, though the required changes for
the $^3\!P_J$ phases are very moderate. 
Also, because in the refit process the pion-nucleon coupling constant was
allowed to change only slightly\cite{bob}, the refitted potential fails
to reproduce the modified phases below 10 MeV for all three $^3\!P_J$ phases.
In other words, the changes in the phases we aimed for could be achieved
only partially, at the price of unwanted changes.
The $\chi^2$ per phase-shift datum of Table \ref{tab6a} for the refitted 
potential is 23, mainly because of the bad description of
the modified $^3\!P_J$ phases below 10 MeV and of $^3\!F_2$ at the higher 
energies. Nevertheless, the $\chi^2$ per
np datum did increase only by about 3\% for the refitted potential
compared to the original potential. This reflects the fact that the $^3\!F_2$
phase is very small and therefore has not much influence on the np data.

Furthermore the modified potential led to very disappointing results in 
the nd
system\cite{ich}. Only an improvement of about 3\% in $A_y$ was achieved at
$E_{lab}=3$ MeV, instead of the necessary 30\%. Similarly, another study
contained in \cite{report}, where the Nijmegen $^3\!P_J$ phases where changed by
up to 3\%, showed that within this restriction a solution of the $A_y$ problem
is not possible. From this we conclude that even if a modification of the
low-energy $^3\!P_J$ phase shifts could be justified and a fit of the potential 
to those modified phases would be possible (which it is not, as shown above) it 
would not solve the $A_y$ problem.

\section{Charge-Independence and Charge-Symmetry Breaking}
\label{cibcsb}

The first of the two assumptions mentioned in the previous section,
namely that there is no CIB and CSB in the $^3\!P_J$ waves,
might be questioned because it was shown in \cite{aypaper}
that the introduction of a very strong CIB and CSB in the Bonn B
NN potential makes it
possible to keep the 2-body observables unchanged while increasing the nd $A_y$
by the necessary amount. However, the $^3\!P_J$ phases were strongly changed.
The $^3\!P_0$ pp phases, for example, were changed by about 15\% at all energies.
This is in clear contradiction with that Nijmegen phase, which has a
statistical uncertainty below 1\% at the energies considered here.
Moreover, two very recent
studies\cite{ruprecht} and \cite{ruprecht1} show that the CIB and CSB used in
\cite{aypaper} cannot by justified on physical grounds. In\cite{ruprecht} and
\cite{ruprecht1} the authors study those CIB and CSB effects that are possible
within the conventional meson-exchange model of Ref. \cite{MHE87}. 
In the meson-exchange picture CIB
and CSB are primarily caused by the differences between the neutral- and
charged-meson masses, as well as the different nucleon masses. In Tables
\ref{tab7} and \ref{tab8} we compare the CIB and CSB as calculated in
\cite{ruprecht} and \cite{ruprecht1} with the one used in \cite{aypaper}
(CIB and CSB effects for the preliminary CD-Bonn99\cite{cdbonn99} are
shown as well for later use). For
$^3\!P_0$ the CIB and CSB used in \cite{aypaper} is not only much stronger than
can be explained by the meson-exchange picture (the CSB is a factor of 20 too
strong), but both also have the wrong sign. The same is true for CSB in the
$^3\!P_1$ partial wave, whereas for the CIB in the 
$^3\!P_1$ and $^3\!P_2$ waves we note the far larger effects for 
\cite{ruprecht} and CD-Bonn99, which has the opposite sign to the modified
Bonn B.

The CIB and CSB as calculated in \cite{ruprecht} and \cite{ruprecht1} have been
built into a new version of the CD-Bonn potential, the so-called CD-Bonn99
\cite{cdbonn99}. In addition, the CD-Bonn99 includes CIB effects from 
irreducible $\pi-\gamma$ exchange as calculated by van Kolck 
{\it et al.}\cite{kolck97}.
The CD-Bonn99 potential has the pion-nucleon coupling constant of
the Nijmegen PSA, $g^2_\pi/4\pi=13.6$, whereas the studies \cite{ruprecht} and
\cite{ruprecht1} (as well as Bonn B)
use a larger pion-nucleon coupling constant of 
$g^2_\pi/4\pi=14.4$.
Thus the CIB and CSB effects in CD-Bonn99 are generally smaller
than those given in \cite{ruprecht} and \cite{ruprecht1} (see Tables
\ref{tab7} and \ref{tab8}). We
had a preliminary version\cite{cdbonn99} of this potential at our disposal,
which we tested in the 3N analyzing powers.
For simplicity we restricted CIB and CSB to the
partial waves that are essential for our problem (i.e., $^1\!S_0$ and
$^3\!P_J$). In all other partial waves we used the np force only. This
calculation is to be compared to one where the CD-Bonn99 np force is used in 
all partial waves. The CIB and CSB built into CD-Bonn99 gives an increase
in the maximum of $A_y$ at $E_{lab}=3$ MeV
of about 4\% and in $i T_{11}$ of about 10\%.
The increase in $A_y$ is far too small to come close to the nd data.
For $i T_{11}$ there are no nd data, but a comparison of a pd calculation
using the AV18 potential with pd data at 3 MeV 
\cite{pisa3nf2} shows a 50\% discrepancy.
Thus we can conclude that although the CIB and CSB effects as built into the
CD-Bonn99 potential go in the correct direction, they are much too small
to explain the discrepancies in the vector analyzing powers.

At first sight it might be surprising that the CD-Bonn99 potential
gives an increase in the analyzing powers (as does the modified Bonn B),
because all CIB effects of the two potentials have opposite signs.
But a closer look at Table \ref{tab7} shows that there is no
inconsistency. Let us remember first that in order
to increase $A_y$ one has to decrease $\delta_{^3\!P_0}$ and increase 
$\delta_{^3\!P_1}$ and $\delta_{^3\!P_2}$ (that is the magnitude of those 
phases; $\delta_{^3\!P_1}$ has a negative sign). We also remember that in a 
charge-dependent Faddeev calculation the CIB effect can be taken into account 
via an effective t-matrix (or with the potential {\it mutatis mutandis})
$t_{eff}=1/3\ t_{np}+2/3\ t_{nn(pp)}$, if we neglect isospin $T=3/2$ channels
($T$ representing the total 3-body isospin).
Thus if we want to get an increase in $A_y$ in a charge-dependent calculation
in comparison to a charge-independent calculation (which uses the np force
only), we need a $\delta_{CIB}$ for $^3\!P_0$ and $^3\!P_1$ with positive
sign (again note that $\delta_{^3\!P_1}$ is negative!) and of negative sign
for $^3\!P_2$. So we see from Table \ref{tab7} that the modified Bonn B
of Ref. \cite{aypaper} has the correct CIB in $^3\!P_0$ in order to
increase $A_y$, but the wrong CIB in the other two phases. But the CIB
effects of the modified Bonn B in these other two phases can be neglected
against the huge CIB in $^3\!P_0$. Thus the large increase in $A_y$
of the modified
Bonn B in comparison to the original Bonn B comes only from $^3\!P_0$.
For CD-Bonn99 on the other hand we  find the correct CIB for an increase of 
$A_y$ for
$^3\!P_1$ and $^3\!P_2$ and the wrong CIB for $^3\!P_0$. Also the CIB
effects in all three phases are of the same order of magnitude. So for
CD-Bonn99 we have an interference of opposite effects, and the
increasing effects just overcome the decreasing one.

We also note that the $A_y$ problem exists in pd scattering\cite{pisa3nf1}, 
as well, and this
involves the well-known pp interaction rather than the poorly known nn force.
Thus for all of these reasons we can exclude CIB and CSB as a solution of the
$A_y$ problem.

\section{Summary and Conclusions}
\label{sum}

In section \ref{Op} we laid down the principles and options of our study
of the $A_y$ puzzle. Any NN potential should fit the NN data
reasonably well, and if it does so, it gives the same answer for $A_y$
as all other potentials. The NN potential has an OPEP as the long-range
part, which is very well known. There is, however, much more uncertainty in the
short-range parts of the NN potential. We argue that CD must be a small effect.
Current 3NF models do not help in $A_y$.

In order to study the possibilities of changes in the NN force we chose the
AV18 model, which is introduced in section \ref{NN}. It consists of the
OPEP and a phenomenological short-range part. This potential has 5 different 
operators that contribute to the $^3\!P_J$ waves (which are
the only important ones for $A_y$).

In section \ref{change} we showed that it is possible to improve the description
of the 3-body $A_y$ and at the same time keep changes in the 2-body $A_2$
small, but that huge changes (at least in the tensor force) are
necessary in order to achieve this. As pointed out in section \ref{size}
such huge changes in the NN potentials can be ruled out, and there is only
very little room for changes in the NN potential at all.

Indeed, the tensor force acts largely through OPEP, and that is the reason
why only the tensor force has a significantly different effect in the
2-body and 3-body systems, as shown in section \ref{opep}. Thus the only way to
increase $A_y$ and keep $A_2$ unchanged at the same time is to change OPEP
by 30\% - 50\%. This is impossible.

Moreover, as we see in section \ref{phases}, the additional requirement
of keeping the $^3\!P_J$ phases unchanged, as well, leads to the requirement of
even more drastic changes in the NN potential.

We also comment in this section on Ref. \cite{werner}, where it is claimed
that there is much room for changes in the low-energy $^3\!P_J$ phases.
Unfortunately this is true only if additional constraints are not applied.
The ambiguities for the $^3\!P_J$ phases found in \cite{werner} can be removed
by performing a multi-energy analysis and by including additional physics 
(i.e., OPEP).

Finally we excluded CIB and CSB as a possible explanation of the $A_y$
puzzle in section \ref{cibcsb}. Although we did not comment on the effects of
long-range electromagnetic forces, it was shown in a recent paper\cite{vincent} 
that they have no major effect on $A_y$.

Thus we have eliminated all possibilities for solving the $A_y$ puzzle on
the 2-body level. Therefore we come to the conclusion that the only possible 
solution for the $A_y$ puzzle must be a 3NF. This 3NF must be a term that has 
not yet been taken into account\cite{3bf}. Because of the nature of the
analyzing power as a difference between cross sections with different spin
direction for one of the incoming particles, it must be a spin-dependent
3NF. Likely candidates are spin-orbit-type 3NFs\cite{3bf}. We also note that
there is a similar problem with the $^5$He energy levels, where the $P_{1/2} - 
P_{3/2}$ splitting is 20-30\% too small\cite{he5}. This seems likely to have 
the same origin as the $A_y$ puzzle and, if so, would have the same solution.

Another strong hint that a 3NF is the solution of the $A_y$ puzzle
is the fact that in \cite{werner} \cite{twk} \cite{wittor} only 
energy-dependent changes (changes in shape)
of the $^3\!P_J$ phases are considered as possible solutions and 
energy-independent changes are ruled out. But an energy-dependent change in 
the NN force (which we do not accept as a possibility, see
section \ref{Op}) is very likely equivalent to adding a 3NF (in the 
three-nucleon systems).
This point is also supported by the fact that the attempts\cite{bob} to
fit AV18 to the energy-dependent modified $^3\!P_J$ phases of Table
\ref{tab6a} were not possible with a satisfactory value for $\chi^2$
per phase-shift datum\cite{bob}.

We would like to point out that in order to investigate such 3NFs it is
desirable to have a consistent description of the NN and 3N force.
Otherwise the complicated interplay between inconsistent NN and 3N forces
might lead to wrong conclusions. A consistent description of NN and 3N
forces, such as realized in the Ruhrpot model\cite{Ruhrpot}, also has the
advantage that the 3NF is essentially parameter-free (i.e., all parameters
occurring in the 3NF are already given by the NN force and its fit to the
NN data base).

We mentioned in section \ref{Op} that an off-shell ambiguity is equivalent to
a 3NF. Is this a serious consideration for our problem? In principle it might 
be, but in practice it is not. The depth of the problem is illustrated in
Ref. \cite{poly}, where a theorem
is proven that any Hamiltonian $H_1$ that contains a 3NF 
can be replaced by a Hamiltonian $H_2$ that does not contain a 3NF,
with $H_1$ and $H_2$ giving the same 3-body binding energy and scattering
matrix. In addition, the Hamiltonian $\bar H_1$ ($H_1$ minus the 3NF)
and $H_2$ give the same 2-body binding energy and scattering matrix.
Thus an off-shell ambiguity in the NN force (needed to define $\bar{H_1}$)
is equivalent to {\bf the whole} 3NF.

In practice the problem is much less dramatic. In our experience\cite{deut}
field-theoretic exercises to define potentials such as OPEP suffer from only 
three types of ambiguity: (1) the BW-TMO ambiguity arising from energy
dependence in the force (discussed in detail in Ref. \cite{twopi}); (2) 
$\mu , \nu$ unitary ambiguities due to chiral representation and choice of 
quasipotential (defined and discussed in \cite{3bf}); (3) ``form'' ambiguities, 
where the entire
structure of the quasipotential equation is altered, such as by squaring the 
relativistic Schr\"{o}dinger equation (see Eqs. (103) and (104) of Ref. 
\cite{mec}). In each case, specifying the form of OPEP eliminates the ambiguity
by fiat. Moreover, the Bonn potentials differ from most others in their
$\mu , \nu$ parameters and this makes little difference in the $A_y$ problem, 
as was shown in Ref. \cite{report}, for example.

We succinctly summarize by stating that if OPEP is not dramatically changed and
if long-range electromagnetic forces are unimportant, the remaining short-range
forces cannot fix the $A_y$ problem. Because these forces are proportionate in 
the 2- and 3-nucleon systems up to quite high energies and are fixed by all 
the NN data in this range, they cannot be altered to resolve the puzzle. This 
conclusion is in clear disagreement with the authors of Ref. \cite{wittor}.
Also, unlike the conjecture of Ref. \cite{twk} we find no evidence that prior 
phase-shift analyses are questionable.
In \cite{wittor} the possibility of a 3NF as an explanation for the $A_y$
discrepancy is ruled out by the argument that ``it is very difficult
to imagine 3N forces can account for the observed energy dependence
of the N-d analyzing power puzzle''. Nevertheless, because every other 
solution is eliminated by our study, what remains must be the correct 
answer\cite{holmes}.

\begin{acknowledgements}
This work was performed in part under the auspices of the U.S.
Department of Energy. The work of D.H. was supported in part by the
Deutsche Forschungsgemeinschaft under Project No. Hu 746/1-2.
The numerical calculations have been performed 
on the Cray T90 of the H\"ochstleistungsrechenzentrum in J\"ulich, 
Germany. We thank R.B. Wiringa for providing information about the 
AV18 potential and for further discussions. We would also like to thank 
R. Machleidt for providing a 
preliminary version of the CD-Bonn99 potential, and for several helpful 
conversations. Furthermore, we thank W. Gl\"ockle, G.M. Hale, A. Kievsky, 
J.J. de Swart, W. Tornow and H. Wita\l a for helpful comments, 
and R. Timmermans for providing a figure.
\end{acknowledgements}

\def\ph{\phantom{$-$}}
\def\mi{$-$}
\begin{table}
\begin{tabular}{r|lllr|lllr}
$E_{lab}$ [MeV]&change&$A_2$&$\Delta$&\%&change&$A_2$&$\Delta$&\%\cr
\hline
  1&$1.1*V_C$&.00023024&\ph.00000064&\ph2.85
   & $.9*V_C$&.00021773&\mi.00000061&\mi2.73\cr
 10&         &.012780  &\ph.000297  &\ph2.38
   &         &.012198  &\mi.000285  &\mi2.28\cr
100&         &.42389   &\mi.00555   &\mi1.29
   &         &.43456   &\ph.00512   &\ph1.19\cr
\hline
  1&$1.1*V_T$&.00019673&\mi.00002712&\mi12.12
   & $.9*V_T$&.00024655&\ph.00002270&\ph10.14\cr
 10&         &.011311  &\mi.001172  &\mi9.39
   &         &.013454  &\ph.000971  &\ph7.78\cr
100&         &.42977   &\ph.00033   &\ph0.08
   &         &.42889   &\mi.00055   &\mi0.13\cr
\hline
  1&$1.1*V_{ls}$&.00026198&\ph.00003813&\ph17.03
   & $.9*V_{ls}$&.00018713&\mi.00003672&\mi16.40\cr
 10&            &.014207  &\ph.001724  &\ph13.81
   &            &.010834  &\mi.001649  &\mi13.21\cr
100&            &.42432   &\mi.00512   &\mi1.19
   &            &.42641   &\mi.00303   &\mi0.71\cr
\hline
  1&$1.1*V_{l^2}$&.00021374&\mi.00001011&\mi4.52
   & $.9*V_{l^2}$&.00023491&\ph.00001106&\ph4.94\cr
 10&             &.012012  &\mi.000471  &\mi3.77
   &             &.012999  &\ph.000516  &\ph4.13\cr
100&             &.43212   &\ph.00268   &\ph0.62
   &             &.42585   &\mi.00359   &\mi0.84\cr
\hline
  1&$1.1*V_{(ls)^2}$&.00022467&\ph.00000082&\ph0.37
   & $.9*V_{(ls)^2}$&.00022302&\mi.00000083&\mi0.37\cr
 10&                &.012501  &\ph.000018  &\ph0.14
   &                &.012462  &\mi.000021  &\mi0.17\cr
100&                &.43312   &\ph.00368   &\ph0.86
   &                &.42558   &\mi.00386   &\mi0.90\cr
\end{tabular}
\caption{\label{tab1} Effects of changes of $\pm 10$\% in the various
parts of the NN potential in the $^3\!P_J$ partial waves on the maximum of the
2-body nucleon analyzing power $A_2$. $\Delta$ gives the difference
between $A_2$ for the original AV18 and the changed one, while \% gives
the change of $A_2$ in percent. The values of the maxima of $A_2$ 
for the original AV18 are
.00022385 at 1 MeV, .012483 at 10 MeV and .42944 at 100 MeV.
}
\end{table}

\begin{table}
\begin{tabular}{r|lllr|lllr}
$E_{lab}$ [MeV]&change&$A_y$&$\Delta$&\%&change&$A_y$&$\Delta$&\%\cr
\hline
  3&$1.1*V_C$&.04685&\ph.00167&\ph3.70
   & $.9*V_C$&.04360&\mi.00158&\mi3.50\cr
\hline
  3&$1.1*V_T$&.04242&\mi.00276&\mi6.11
   & $.9*V_T$&.04716&\ph.00198&\ph4.38\cr
\hline
  3&$1.1*V_{ls}$&.05254&\ph.00736&\ph16.29
   & $.9*V_{ls}$&.03829&\mi.00689&\mi15.25\cr
\hline
  3&$1.1*V_{l^2}$&.04287&\mi.00231&\mi5.11
   & $.9*V_{l^2}$&.04775&\ph.00257&\ph5.69\cr
\hline
  3&$1.1*V_{(ls)^2}$&.04482&\mi.00036&\mi0.80
   & $.9*V_{(ls)^2}$&.04555&\ph.00037&\ph0.82\cr
\end{tabular}
\caption{\label{tab2} Same as Table \ref{tab1}, but for the 3-body nucleon 
analyzing power $A_y$. The value of the maximum of $A_y$ at 3 MeV for the
original AV18 is .04518.
}
\end{table}

\begin{table}
\begin{tabular}{rrrrr|rr}
$\delta_C$&$\delta_{l^2}$&$\delta_{(ls)^2}$&$\delta_T$&$\delta_{ls}$&
$\delta_{A_2}$&$\delta_{A_y}$\cr
\hline
10&\mi10&    0&52.0&32.0&3.6&54\cr
10&\mi10&\mi10&49.4&30.4&5.1&55\cr
22&\mi22&    0&46.4&22.4&6.2&55\cr
22&\mi22&\mi22&40.7&18.9&8.5&57\cr
30&\mi30&    0&42.7&16.0&7.4&56\cr
30&\mi30&\mi30&34.9&11.2&9.6&57\cr
\end{tabular}
\caption{\label{tab3} Possible solutions of Eqs. (\ref{eq2}) and (\ref{eq3}).
The effects of the changes to $A_2$ and $A_y$ are given as well.
}
\end{table}

\begin{table}
\begin{tabular}{r|lllr|lllr}
$E_{lab}$ [MeV]&change&$A_2$&$\Delta$&\%&change&$A_2$&$\Delta$&\%\cr
\hline
  1&$1.1*V^{SR}_C$  &.00023159&\ph.00000774&\ph3.46
   &$1.1*V^{OPEP}_C$&.00022256&\mi.00000129&\mi0.58\cr
 10&                &.012856  &\ph.000373  &\ph2.99
   &                &.012410  &\mi.000073  &\mi0.58\cr
100&                &.42174   &\mi.00770   &\mi1.79
   &                &.43150   &\ph.00206   &\ph0.48\cr
\hline
  1&$1.1*V^{SR}_T$  &.00022601&\ph.00000216&\ph0.96
   &$1.1*V^{OPEP}_T$&.00019419&\mi.00002966&\mi13.25\cr
 10&                &.012591  &\ph.000108  &\ph0.87
   &                &.011183  &\mi.001300  &\mi10.41\cr
100&                &.42943   &\mi.00001   &\ph0.
   &                &.42986   &\ph.00042   &\ph0.10\cr
\end{tabular}
\caption{\label{tab4} Same as Table \ref{tab1}, but for the short-range
and OPEP parts of the central and tensor force separately.
}
\end{table}

\begin{table}
\begin{tabular}{r|lllr|lllr}
$E_{lab}$ [MeV]&change&$A_y$&$\Delta$&\%&change&$A_y$&$\Delta$&\%\cr
\hline
  3&$1.1*V^{SR}_C$  &.04732&\ph.00214&\ph4.74
   &$1.1*V^{OPEP}_C$&.04473&\mi.00045&\mi1.00\cr
\hline
  3&$1.1*V^{SR}_T$  &.04552&\ph.00034&\ph0.75
   &$1.1*V^{OPEP}_T$&.04200&\mi.00318&\mi7.04\cr
\end{tabular}
\caption{\label{tab5} Same as Table \ref{tab2}, but for the short-range
and OPEP parts of the central and tensor force separately.
}
\end{table}

\begin{table}
\begin{tabular}{r|lllr|lllr}
phase&change&phase&$\Delta$&\%&change&phase&$\Delta$&\%\cr
\hline
$^3\!P_0$&$1.1*V_C$&\ph.17980&\mi.00065&\mi0.36
       & $.9*V_C$&\ph.18111&\ph.00066&\ph0.37\cr
$^3\!P_1$&         &\mi.10831&\mi.00094&\ph0.88
       &         &\mi.10644&\ph.00093&\mi0.87\cr
$^3\!P_2$&         &\ph.022130&\mi.00040&\mi1.77
       &         &\ph.022951&\ph.00042&\ph1.87\cr
\hline
$^3\!P_0$&$1.1*V_T$&\ph.20495&\ph.02450&\ph13.58
       & $.9*V_T$&\ph.15680&\mi.02365&\mi13.11\cr
$^3\!P_1$&         &\mi.11575&\mi.00838&\ph7.80
       &         &\mi.09891&\ph.00846&\mi7.88\cr
$^3\!P_2$&         &\ph.025649&\ph.00312&\ph13.85
       &         &\ph.019527&\mi.00300&\mi13.33\cr
\hline
$^3\!P_0$&$1.1*V_{ls}$&\ph.17883&\mi.00162&\mi0.90
       & $.9*V_{ls}$&\ph.18217&\ph.00172&\ph0.95\cr
$^3\!P_1$&            &\mi.10768&\mi.00031&\ph0.29
       &            &\mi.10706&\ph.00031&\mi0.29\cr
$^3\!P_2$&            &\ph.024190&\ph.00166&\ph7.37
       &            &\ph.021003&\mi.00153&\mi6.77\cr
\hline
$^3\!P_0$&$1.1*V_{l^2}$&\ph.18042&\mi.00003&\mi0.02
       & $.9*V_{l^2}$&\ph.18049&\ph.00004&\ph0.02\cr
$^3\!P_1$&             &\mi.10735&\ph.00002&\mi0.02
       &             &\mi.10740&\mi.00003&\ph0.03\cr
$^3\!P_2$&             &\ph.021871&\mi.00066&\mi2.92
       &             &\ph.023253&\ph.00072&\ph3.21\cr
\hline
$^3\!P_0$&$1.1*V_{(ls)^2}$&\ph.17921&\mi.00124&\mi0.69
       & $.9*V_{(ls)^2}$&\ph.18173&\ph.00128&\ph0.71\cr
$^3\!P_1$&                &\mi.10751&\mi.00014&\ph0.13
       &                &\mi.10724&\ph.00013&\mi0.12\cr
$^3\!P_2$&                &\ph.022001&\mi.00053&\mi2.34
       &                &\ph.023070&\ph.00054&\ph2.40\cr
\end{tabular}
\caption{\label{tab6} Same as Table \ref{tab1}, but for the np $^3\!P_J$
phases at $E_{lab}=1$ MeV. The values for the original AV18 are .18045 for
$^3\!P_0$, \mi.10737 for $^3\!P_1$, and .022529 for $^3\!P_2$.
}
\end{table}

  \def\turntext#1{\special{ps:currentpoint currentpoint translate
     #1 neg rotate neg exch neg exch translate}}
  \def\Ytext#1{\setbox0=\hbox{#1}\dimen0=\ht0 \advance\dimen0 by\dp0
     \hbox to\dimen0{\hss\vbox to\wd0{\vss
\turntext{90}%
       \hbox to0pt{\hss\vbox to0pt{\vss\box0\vss}\hss}
\vss}\hss}}
  \def\rotatedtext#1,#2{
\turntext{#1}%
       \hbox to0pt{\hss\vbox to0pt{\vss\hbox{#2}\vss}\hss}%
}

\Ytext{\hbox to 22.5truecm{\vbox{
\begin{table}
\begin{tabular}{r|rrrr|rrrr|rrrr|rrrr|rrrr}
$E_{lab}$
&\multicolumn{4}{c|}{$^3\!P_0$}&\multicolumn{4}{c|}{$^3\!P_1$}
&\multicolumn{4}{c|}{$^3\!P_2$}&\multicolumn{4}{c|}{$\epsilon_2$}
&\multicolumn{4}{c}{$^3\!F_2$}\\
&AV18&mod.&ach.&\%&AV18&mod.&ach.&\%&AV18&mod.&ach.&\%
&AV18&mod.&ach.&\%&AV18&mod.&ach.&\%\\
\hline
  1.&  0.18&  0.171&  0.178& 4.1& -0.11& -0.105& -0.107& 1.9
& 0.02& 0.024& 0.023&-4.3&-0.00&-0.001&-0.001&  0.&0.00&0.000&0.000&  0.\\
  5.&  1.64&  1.581&  1.619& 2.4& -0.93& -0.913& -0.921& 0.9
& 0.26& 0.267& 0.262&-1.9&-0.05&-0.049&-0.049&  0.&0.00&0.002&0.002&  0.\\
 10.&  3.71&  3.616&  3.649& 0.9& -2.04& -2.021& -2.026& 0.2
& 0.72& 0.738& 0.733&-0.7&-0.19&-0.185&-0.184&-0.5&0.01&0.011&0.011&  0.\\
 25.&  8.32&  8.320&  8.167&-1.9& -4.82& -4.819& -4.778&-0.9
& 2.57& 2.570& 2.601& 1.2&-0.77&-0.768&-0.762&-0.8&0.08&0.086&0.083&-3.6\\
 50.& 10.99& 10.993& 10.801&-1.8& -8.15& -8.145& -8.079&-0.8
& 5.86& 5.863& 5.895& 0.5&-1.68&-1.678&-1.657&-1.3&0.28&0.280&0.259&-8.1\\
100.&  8.69&  8.691&  8.595&-1.1&-13.07&-13.065&-12.998&-0.5
&11.00&10.998&10.980&-1.6&-2.69&-2.692&-2.653&-1.5&0.67&0.668&0.590&-13.2\\
150.&  3.78&  3.780&  3.788& 2.1&-17.28&-17.284&-17.238&-0.3
&14.12&14.120&14.065&-0.4&-2.95&-2.946&-2.910&-1.2&0.98&0.979&0.863&-13.4\\
200.& -1.43& -1.426& -1.350&-5.6&-21.22&-21.221&-21.189&-0.2
&15.86&15.862&15.799&-0.4&-2.82&-2.822&-2.792&-1.1&1.15&1.149&1.030&-11.6\\
250.& -6.41& -6.410& -6.299&-1.8&-24.95&-24.953&-24.921&-0.1
&16.70&16.694&16.640&-0.3&-2.54&-2.539&-2.503&-1.4&1.10&1.097&0.997&-10.0\\
300.&-11.06&-11.056&-10.935&-1.1&-28.49&-28.495&-28.450&-0.2
&16.91&16.908&16.863&-0.3&-2.21&-2.207&-2.149&-2.7&0.77&0.766&0.688&-11.3\\
350.&-15.36&-15.358&-15.245&-0.7&-31.85&-31.851&-31.783&-0.2
&16.69&16.686&16.646&-0.2&-1.88&-1.879&-1.786&5.2&0.14&0.137&0.062&-21.0
\end{tabular}
\caption{\label{tab6a} Phase shift parameters of the original AV18 potential 
\protect\cite{AV18} compared to phases which were modified in the spirit
of \protect\cite{werner} in order to give an increased prediction for $A_y$.
Also given are the phases which could be achieved by the potential refitted
to the modified phases \protect\cite{bob}. Note that the pion-nucleon coupling 
constant was only allowed to change slightly during the refit process. The 
difference in percent between the modified and the achieved phases is given as 
well.}
\end{table}
}}}

\newpage
\begin{table}
\begin{tabular}{r|rrr|rrr|rrr}
&\multicolumn{3}{c|}{$^3\!P_0$}&\multicolumn{3}{c|}{$^3\!P_1$}
&\multicolumn{3}{c}{$^3\!P_2$}\\
$E_{lab}$ [MeV]
&$\delta_{CIB}^{CDB99}$&$\delta_{CIB}^{\protect\cite{ruprecht}}$
&$\delta_{CIB}^{mod.\ BB}$
&$\delta_{CIB}^{CDB99}$&$\delta_{CIB}^{\protect\cite{ruprecht}}$
&$\delta_{CIB}^{mod.\ BB}$
&$\delta_{CIB}^{CDB99}$&$\delta_{CIB}^{\protect\cite{ruprecht}}$
&$\delta_{CIB}^{mod.\ BB}$\\
\hline
5&     \mi.237&\mi.250& .515&.112&.117&0.     &\mi.011&\mi.011&0.\\
10(12)&\mi.466&\mi.492&1.405&.198&.206&\mi.015&\mi.032&\mi.032&.005\\
25&    \mi.822&\mi.858&2.575&.314&.322&\mi.025&\mi.103&\mi.101&.015\\
50&    \mi.943&\mi.960&3.275&.368&.366&\mi.035&\mi.188&\mi.184&.045\\
\end{tabular}
\caption{\label{tab7} CIB effects in degrees for the preliminary
CD-Bonn99 potential\protect\cite{cdbonn99} and
from\protect\cite{ruprecht} compared to CIB in the
modified Bonn B of\protect\cite{aypaper}. $E_{lab}=10$ MeV refers to 
CD-Bonn99 and \protect\cite{ruprecht},
and 12 MeV to the modified Bonn B. The CIB effect is defined as
$\delta_{CIB}\equiv \delta_{np}-.5*(\delta_{nn} +\delta_{pp})$.}
\end{table}

\begin{table}
\begin{tabular}{r|rrr|rrr|rrr}
&\multicolumn{3}{c|}{$^3\!P_0$}&\multicolumn{3}{c|}{$^3\!P_1$}
&\multicolumn{3}{c}{$^3\!P_2$}\\
$E_{lab}$ [MeV]
&$\delta_{CSB}^{CDB99}$&$\delta_{CSB}^{\protect\cite{ruprecht1}}$
&$\delta_{CSB}^{mod.\ BB}$
&$\delta_{CSB}^{CDB99}$&$\delta_{CSB}^{\protect\cite{ruprecht1}}$
&$\delta_{CSB}^{mod.\ BB}$
&$\delta_{CSB}^{CDB99}$&$\delta_{CSB}^{\protect\cite{ruprecht1}}$
&$\delta_{CSB}^{mod.\ BB}$\\
\hline
5&     .008&.009&\mi .23&\mi.002&\mi.002&\mi.02&.002&.003&0.\\
10(12)&.018&.019&\mi .63&\mi.003&\mi.002&\mi.07&.006&.007&.01\\
25&    .040&.042&\mi1.15&\mi.001&     0.&\mi.13&.022&.025&.03\\
50&    .056&.057&\mi1.49&   .006&   .010&\mi.11&.051&.056&.09\\
\end{tabular}
\caption{\label{tab8} CSB effects in degrees for the preliminary
CD-Bonn99 potential\protect\cite{cdbonn99} and
from\protect\cite{ruprecht1} compared to CSB in the
modified Bonn B of\protect\cite{aypaper}. $E_{lab}=10$ MeV refers to 
CD-Bonn99 and \protect\cite{ruprecht1},
and 12 MeV to the modified Bonn B. The CSB effect is defined as
$\delta_{CSB}\equiv \delta_{nn} -\delta_{pp}$.}
\end{table}

\input epsf

\input{prepictex}
\input{pictex}
\input{postpictex}

\begin{figure}
\input{fig0.tex}
\caption{\label{fig0} The $^3\!P_0$ phase shift divided by $E^{3/2}$.
The solid line denotes AV18, the dashed line AV18 with $V_T^{SR}$ set to 
zero, while the dotted line is the difference between the dashed and solid 
lines.}
\end{figure}
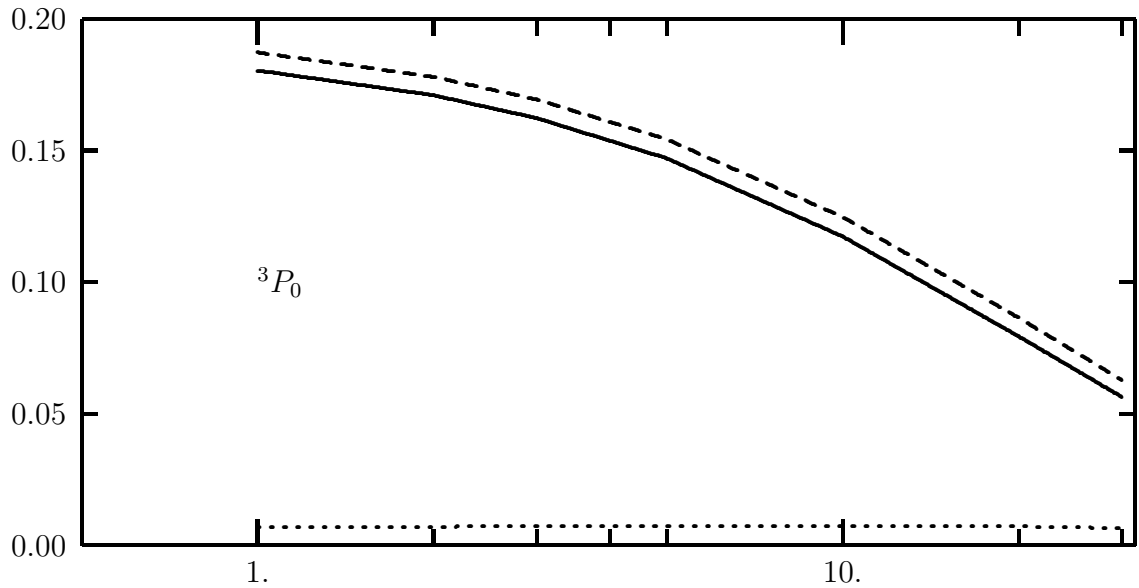

\begin{figure}
\input{fig1.tex}
\caption{\label{fig1} Same as Figure \ref{fig0} but for $^3\!P_1$.}
\end{figure}

\begin{figure}
\input{fig2.tex}
\caption{\label{fig2} Same as Figure \ref{fig0} but for $^3\!P_2$.}
\end{figure}
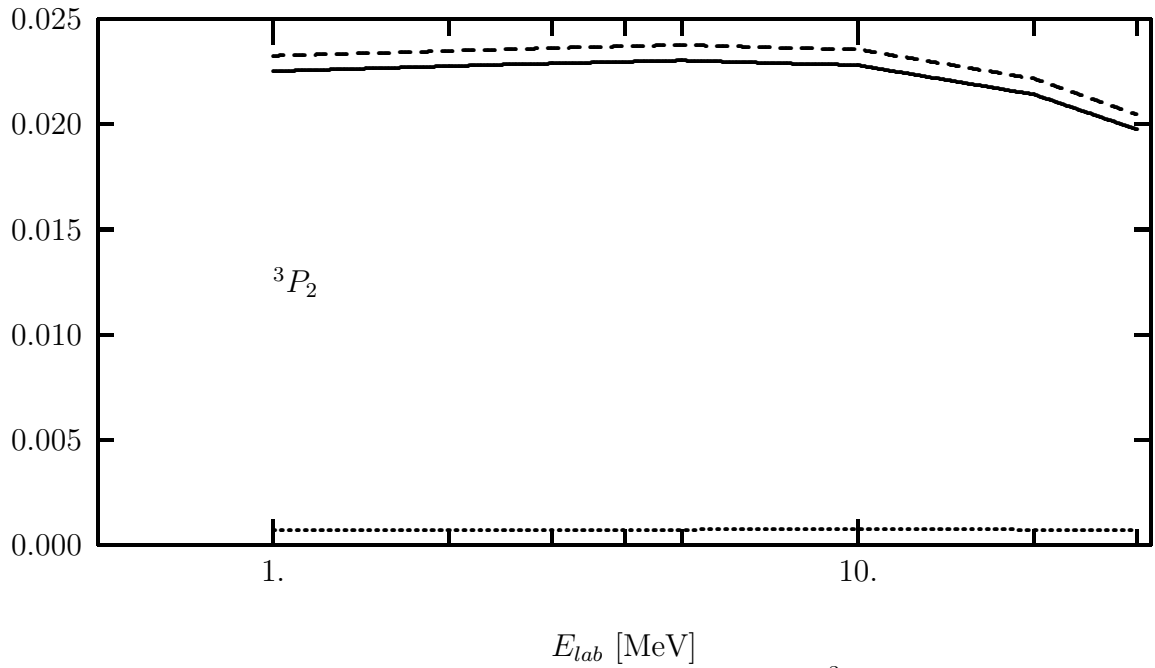

\end{document}

%% file: fig0.tex
 
  \beginpicture
\setcoordinatesystem units < 7.7778 true cm, 35. true cm>
\setplotarea x from -.3 to  1.5, y from .0 to .2
  \linethickness 1.2 pt
  \setplotsymbol ({\large .})
\axis bottom label {$E_{lab}$ [MeV]} ticks logged in numbered at 1. 10. /
unlabeled short in at 2. 3. 4. 5. 20. 30. / /
\axis top ticks logged in unlabeled at 1. 10. /
unlabeled short in at 2. 3. 4. 5. 20. 30. / /
\axis left ticks short in numbered from .00 to .20 by .05 /
\axis right ticks short in unlabeled from .00 to .20 by .05 /

\put {$^3P_0$} [l] at 0. .1

\plot
0.      .18045
.30103  .17102
.47712  .16232
.69897  .14709
1.      .11737
1.30103 .07938
1.47712 .05640
/

\setdashpattern <4pt, 4pt>
\plot
0.      .18730
.30103  .17799
.47712  .16945
.69897  .15432
1.      .12477
1.30103 .08646
1.47712 .06285
/

\setdashpattern <1pt, 4pt>
\plot
0.      .00685
.30103  .00697
.47712  .00713
.69897  .00723
1.      .00740
1.30103 .00708
1.47712 .00645
/

 \endpicture

%% file: fig1.tex
 
  \beginpicture
\setcoordinatesystem units < 7.7778 true cm, 63.636 true cm>
\setplotarea x from -.3 to  1.5, y from -.11 to .0
  \linethickness 1.2 pt
  \setplotsymbol ({\large .})
\axis bottom label {$E_{lab}$ [MeV]} ticks logged in numbered at 1. 10. /
unlabeled short in at 2. 3. 4. 5. 20. 30. / /
\axis top ticks logged in unlabeled at 1. 10. /
unlabeled short in at 2. 3. 4. 5. 20. 30. / /
\axis left ticks short in numbered from -.10 to .00 by .02 /
\axis right ticks short in unlabeled from -.10 to .00 by .02 /

\put {$^3P_1$} [l] at 0. -.05

\plot
0.      -.10737
.30103  -.10006
.47712  -.09366
.69897  -.08302
1.      -.06459
1.30103 -.04459
1.47712 -.03399
/

\setdashpattern <4pt, 4pt>
\plot
0.      -.10834
.30103  -.10103
.47712  -.09464
.69897  -.08401
1.      -.06560
1.30103 -.04564
1.47712 -.03504
/

\setdashpattern <1pt, 2pt>
\plot
0.      -.00097
.30103  -.00097
.47712  -.00098
.69897  -.00099
1.      -.00101
1.30103 -.00105
1.47712 -.00105
/

 \endpicture

%% file: fig2.tex
 
  \beginpicture
\setcoordinatesystem units < 7.7778 true cm, 280. true cm>
\setplotarea x from -.3 to  1.5, y from .0 to .025
  \linethickness 1.2 pt
  \setplotsymbol ({\large .})
\axis bottom label {$E_{lab}$ [MeV]} ticks logged in numbered at 1. 10. /
unlabeled short in at 2. 3. 4. 5. 20. 30. / /
\axis top ticks logged in unlabeled at 1. 10. /
unlabeled short in at 2. 3. 4. 5. 20. 30. / /
\axis left ticks short in numbered from .000 to .025 by .005 /
\axis right ticks short in unlabeled from .000 to .025 by .005 /

\put {$^3P_2$} [l] at 0. .0125

\plot
0.      .022529
.30103  .022747
.47712  .022894
.69897  .023029
1.      .022802
1.30103 .021413
1.47712 .019739
/

\setdashpattern <4pt, 4pt>
\plot
0.      .023251
.30103  .023472
.47712  .023623
.69897  .023762
1.      .023544
1.30103 .022151
1.47712 .020453
/

\setdashpattern <1pt, 2pt>
\plot
0.      .000722
.30103  .000725
.47712  .000729
.69897  .000733
1.      .000742
1.30103 .000738
1.47712 .000714
/

 \endpicture